%% file: adamo_bastian_final2.tex
\documentclass[graybox]{svmult}

\usepackage{mathptmx}       
\usepackage{helvet}         
\usepackage{courier}        
\usepackage{type1cm}        
\usepackage{makeidx}         
\usepackage{graphicx}        
\usepackage{multicol}        
\usepackage[bottom]{footmisc}

\def\aap{A\&A}                
\def\aj{AJ}                   
\def\apj{ApJ}                 
\def\apjl{ApJ}                
\def\apjs{ApJS}               
\def\araa{ARA\&A}             

\def\mnras{MNRAS}%

\def\gamma{\hbox{$\Gamma$}}

\def\msun{\hbox{M$_\odot$}}
\def\mstar{\hbox{M$_\star$}}
\def\t4{\hbox{t$_{\rm 4}$}}

\def\dndt{\hbox{${d{\rm N/}}{dt}$}}

\makeindex             

\begin{document}

\title*{The Lifecycle of Clusters in Galaxies}
\author{Angela Adamo and Nate Bastian\\{\em To appear in The Origin of Stellar Clusters, 2016, ed. S. Stahler (Springer)}}
\authorrunning{Adamo \& Bastian}
\institute{Angela Adamo \at The Oskar Klein Centre, Department of Astronomy, AlbaNova, Stockholm University, SE-106 91 Stockholm, Sweden \email{adamo@astro.su.se}
\and Nate Bastian \at Astrophysics Research Institute, Liverpool John Moores University, 146 Brownlow Hill, Liverpool L3 5RF, UK, \email{N.J.Bastian@ljmu.ac.uk}}
%
%
\maketitle
\vspace{-3.5cm}
\abstract{We review many of the basic properties of star cluster systems, and focus in particular on how they relate to their host galaxy properties and ambient environment.  The cluster mass and luminosity functions are well approximated by power-laws of the form $Ndm \propto M^{\alpha}dm$, with $\alpha\sim-2$ over most of the observable range.  However, there is now clear evidence that both become steeper at high masses/luminosities, with the value of the downward turn dependent on environment.  The host galaxy properties also appear to affect the cluster formation efficiency ($\Gamma$ - i.e., the fraction of stars that form in bound clusters), with higher star-formation rate density galaxies having higher $\Gamma$ values.  Within individual galaxies, there is evidence for $\Gamma$ to vary by a factor of $3-4$, likely following the molecular gas surface density, in agreement with recent predictions.  Finally, we discuss cluster disruption and its effect on the observed properties of a population, focussing on the age distribution of clusters.  We briefly discuss the expectations of theoretical and numerical studies, and also the observed distributions in a number of galaxies.  Most observational studies now find agreement with theoretical expectations, namely nearly a constant cluster age distribution for ages up to $\sim100$~Myr (i.e. little disruption), and a drastic steepening above this value caused by a combination of cluster disruption and incompleteness.  Rapid cluster disruption for clusters with ages $<100$~Myr is ruled out for most galaxies.}

\section{Introduction}
\label{sec:intro}

Galactic and extragalactic star-forming regions show that the vast majority of stars are formed in clustered environments, i.e. in the densest cores of giant molecular clouds (GMCs). Clustering is a common feature observed in local star-forming regions, caused by the fractal properties of the ISM under the effect of turbulence (Elmegreen 
\& Efremov~1997). As result, star formation appears to be a hierarchical process, with GMC complexes on large scales ($\sim$1 kpc), and young star clusters (YSCs) at the bottom of the hierarchy forming the densest and only bound structures (Elmegreen~2011, Hopkins~2013a). Turbulence is one of the driving mechanisms which governs star formation. Because turbulence is a scale-free process,  both gas and stars follow continuum density distributions that are described by lognormal functions. Stars will form only in regions which have gas densities above a certain threshold (Kainulainen et al.~2014), and only a fraction of these stars will be formed in systems dense enough to be  gravitationally bound (Bressert et al.~2010).  Throughout this chapter we will focus on YSCs that are gravitationally bound, i.e. systems that are older than a dynamical time, which separates bound clusters from unbound associations (e.g., Gieles \& Portegies Zwart~2011). We will also only address properties of clusters with ages less than a few hundred Myr.  YSCs typically contain $10^2 - 10^7$ stars, and have effective radii between  $1-10$ pc, often leading to systems with densities exceeding that observed in globular clusters (Portegies Zwart et 
al.~2010). 

YSCs are easily detected with the resolving power of the Hubble Space Telescope (HST), in star-forming galaxies as distant as $\sim$100 Mpc (e.g. Adamo et al.~2010a, Fedotov et al.~2011) and many may remain bound for billions of years. Hence they can keep records of the star formation history (SFH) of their host galaxy. Indeed, globular clusters (GCs), remnants of the extreme star formation process that occurred in a much younger Universe, are likely the ancient counterparts of the YSCs we observe in local galaxies (e.g.,~Kruijssen 2014). 
In this contribution we will focus on the statistical and physical properties which characterise YSCs and their relation to star-formation more generally. In particular we will discuss how the galactic environment of the parent galaxy influences the YSC population within it. 
\begin{svgraybox}
Potentially, YSCs can bridge the divide between the sub-pc scales of star-formation and the kpc scales of galaxy formation and evolution.  They can be used as tracers of star formation in space and time, provided that we have a full understanding of their formation, evolution, and disruption as a function of the galactic environment.
\end{svgraybox}

\section{Cluster populations}
\label{sec:cluster_populations}

While much can be learned by studying individual clusters in exquisite detail, many works have focussed on entire cluster populations to see 1) the full range of properties that clusters can have and their statistical distributions, and 2) how these distributions relate to each other, 3) how the host environment affects the initial distributions and how they  evolve with time.

Photometry can be used to estimate the age, mass and extinction of a cluster by comparing the observed cluster luminosity and colours to simple stellar population (SSP) models (where all stars have the same age and metallicity within some small tolerance).  Most studies to date have focussed on the UV and optical parts of the spectrum, where the changes in the overall spectral energy distribution of the cluster change most rapidly as a function of age (although see Gazak et al.~2013 for a near-IR photometric age indicator).  Hence, by obtaining imaging in the U, B, V, and I bands, and including a narrow band filter like H$\alpha$ to break the age-extinction degeneracy, we can estimate the basic parameters of tens or hundreds of clusters at once (c.f., Anders et al.~2004).  Alternatively, UV and optical spectroscopy of massive clusters can be used to infer more accurate ages, and hence masses and extinctions, along with estimates of the cluster radial velocity and metallicity. However, this only allows for the study of single (or tens, with multi-slit observations) clusters, making large samples prohibitively expensive to obtain (e.g., Trancho et al.~2007; Konstantopoulos et al.~2009).  One caveat, however, to these types of studies, is that by using traditional SSP models, an implicit assumption is made that the initial mass function of stars within each cluster is fully sampled. However, this is only strictly valid for the most massive clusters $>10^5-10^6$~\msun. For lower mass clusters, stochastic sampling of the IMF can have dramatic affects on the estimated ages, masses, and extinctions (e.g., Fouesneau \& Lan{\c c}on~2010), or even whether or not a cluster is detected (Silva-Villa \& Larsen~2011).  As such, care must be taken when interpreting the results for lower mass clusters. Often a lower mass limit of $5000$~\msun\ is adopted\footnote{Although stochastic effects are still present to some level at this mass for young ages.}.  Additionally, throughout this chapter, and for most studies in the literature, it is assumed that clusters are well approximated as an SSP (i.e., they have negligible spreads in age and abundance within them), which appears to be good approximation (e.g., Longmore et al.~2014).

\subsection{Cluster formation}
In this section we will provide a statistical description of the main YSC population properties and how they are intrinsically linked to star formation more generally and to the properties of their parent galaxies. The interested reader can find an excellent review of the most recent theories and observational evidence on cluster formation  in the work by Longmore et al.~(2014). 
    
\subsubsection{The Cluster Mass and Luminosity Functions}
\label{sec:icmf}

During the past two decades, numerous observational studies have provided clear evidence that the initial cluster mass function (ICMF) can be well described by a power-law distribution $dN/dM \sim M^{\alpha}$, with index $\alpha \sim -2$ (e.g. Zhang \& Fall~1999,  Bik et al.~2003, Hunter et al.~2003, de Grijs et al.~2003). This same distribution is also found for the youngest (i.e. embedded) clusters/assocations (e.g., Lada \& Lada~2003). The index of the ICMF can be understood in the framework of the hierarchical properties of the ISM, which makes star formation a scale-free process due to supersonic motions in the presence of turbulence and self-gravity (Elmegreen~2006, Hopkins~2013b) .  For this reason, the high mass end of the stellar IMF, most of the cluster mass range, and upper end of the GMC mass functions are reasonably approximated by power-laws, with similar indices ($-2\pm0.3$, Kennicutt \& Evans~2012). 

The ICMF appears to be sampled stochastically within galaxies, so it is desirable to observe a large and massive cluster population in merging galaxies with high star-formation rates (see Section~\ref{cfe}). However, when we look at cluster formation in dwarf galaxies, the change can be quite drastic. In these systems, star  and cluster formation is a sporadic event, and during peaks of star formation, dwarf galaxies can form very massive clusters or potentially, few or no clusters (Billett et al.~2002, Cook et al.~2012). In spiral galaxies, on the other hand, star formation is largely constant over a large time range. In these systems, cluster populations are often continuous in their age and mass distributions.

However, the mass range over which the power law has been fitted varies from study to study, hence a direct comparison between galaxies has been somewhat limited.  Nevertheless, from recent studies, it is becoming increasingly clear that the ICMF of some galaxies has a turn-down at high masses, the exact location of which varies from galaxy to galaxy, and even within a single galaxy (Larsen~2009, Bastian et al.~2012). The Antennae merger system, for example, has a power-law ICMF with index close to $-2$ within a mass range from $10^4$ to $10^6$ \msun (Zhang \& Fall~1999), with any turn-down being above $10^6$~\msun (Portegies Zwart et al.~2010).  It is interesting to notice that, in spiral galaxies, YSC masses rarely reach the range typically observed in merger systems, although there are some exceptions (e.g. NGC\,6946, Larsen et al.~2001). 

However, Larsen (2009) showed that the ICMF of the Milky Way cannot be reconciled with a power law function within the same mass range as for the Antennae, namely $10^2$ to $10^7$~\msun. It is more likely that the upper mass end of the ICMF of the Milky Way is closer to $\sim10^5$~\msun. This value is not a sharp truncation, but the probability that a cluster can form with a mass significantly larger than this value rapidly approaches zero. A Schechter~(1976) function,  
\begin{equation}
\frac{dN}{dM} \propto (M/\mstar)^{\alpha} exp(M/\mstar), 
\end{equation}
is a valid approximation of this distribution because it can describe, simultaneously, the power-law distribution with index $\alpha$ (generally taken to be $-2$) for clusters with masses below a characteristic mass, \mstar, and an exponential distribution for higher masses.  Gieles et al.~(2006), Gieles~(2009) and Larsen (2009) have shown that 
a Schechter function  is a better approximation of the high mass cluster distribution than a pure power-law function for a sample of dwarf and spiral galaxies. 

The characteristic mass, \mstar, appears to vary as function of galactic environment. Larsen (2009) suggested that  spirals have \mstar$\sim 1-2 \times 10^5$~\msun, while the Antennae has most likely a higher truncation mass (\mstar $\sim 10^6$~\msun). The presence of an upper mass limit or a truncation mass in the ICMF suggests that the host galaxies will unlikely be able to produce clusters with masses, $M \gg$\mstar. However, it is important to bear in mind, that cluster formation is a stochastic process and that the ICMF is stochastically populated. The truncation mass is only a value above which it becomes unlikely (but not impossible) to form clusters. 

The presence of such an upper limit in the ICMF could be linked to the ability of the galaxy to form massive GMCs. It is known that shear and streaming motions in spiral systems destroy GMCs, while in environments like the Antennae, the external pressure exerted on the gas makes it possible to form very massive GMCs and GMC complexes. Since clusters form in GMCs (and must have masses less than their progenitor GMCs) the difference in GMC masses observed, for example, in the Milky Way and in the Antennae may explain why the Milky Way is unlikely to form clusters more massive than a few times $10^5$~\msun (Larsen~2009). A recent high-spatial resolution study of the GMC population in the grand-design spiral M\,51 has revealed how GMC properties change as function of the galactic environment (Colombo et al.~2014). In particular, the maximum mass of the GMCs is tightly related to the dynamical environment of M\,51, with higher masses found in the central regions and spiral arms and less massive ones in the inter-arm regions. Kruijssen~(2014), using both theoretical arguments and observations, proposed that the maximum GMC mass is linked to the Toomre mass and therefore to the gas surface density within the region. The Toomre mass is also a fairly good prediction of the characteristic ICMF mass, \mstar, assuming star formation and cluster formation efficiency are known. 

In support of the environmental dependency of the truncation mass of the ICMF, Bastian et al.~(2012)  found a different truncation mass of the cluster population in an inner and outer region (M$_{\star}^{in} \sim 1.6 \times 10^5$ and M$_{\star}^{out} \sim 0.5 \times 10^5$~\msun) of another grand-design spiral galaxy, M\,83. Similar results have been found for NGC~4041(Konstantopoulos et al.~2013).  The difference of the truncation mass in the inner and outer field  can be explained by the difference in the gas surface density within the two regions. Using the same data as Bastian et al.~(2012), Chandar et al.~(2010, 2014) reported that the mass functions of the cluster population in these two regions follow a pure power-law distribution, with index $-2$, in the inner region, but is significantly steeper (over a similar mass range) in the outer region.  When approximating an ICMF as a single power-law, this is the type of behaviour expected if a truncation is present.  Hence, the two studies appear to be consistent, finding evidence of a truncation (or at least a steepening) at high masses.

 Larsen~(2006) has shown that the number of clusters populating the high mass bins is small and it is usually dominated by the size of the cluster population. If the truncation mass is about $10^4$~\msun\ then a cluster population of a few hundred clusters could be enough to statistically distinguish between a pure power law ICMF without upper limits and a Schechter ICMF. An order of magnitude higher truncation mass ($\sim10^5$~\msun) requires a much more numerous cluster population (a factor of 10 higher) to populate significantly the high mass bins. Therefore it is statistically challenging to trace an upper mass truncation in local galaxies and large cluster populations are needed if standard histograms are used.  Instead, cumulative distributions or statistics that use just the brightest/most massive clusters do a better job at finding whether a truncation is present, in the limit of relatively small cluster populations (Ma{\'{\i}}z Apell{\'a}niz \& {\'U}beda~2005, Maschberger \& Kroupa~2009).

It is worth mentioning that the globular cluster mass function is also better fitted by an evolved Schechter function (it takes into account the effect of the temporal evolution of cluster masses, Jord{\'a}n et al.~2007). These authors also found that the truncation mass of the globular cluster mass function is positively correlated to the total B band luminosity (stellar mass) of the host galaxy. Dynamical friction cannot alone explain the observed trend, therefore it must be linked to the physical properties of the galaxy at the moment a significant fraction of their globular cluster population was formed (Kruijssen~2015). 

While the ICMF is the underlying physical distribution that we wish to understand, observational works often focus on the cluster luminosity function (CLF), as this does not require one to estimate the age of each cluster (a necessary step in order to apply the age dependent mass-to-light ratio from SSP models).  As for the ICMF, most studies have found that the CLF is well approximated by a power-law with an index of $\sim-2$ over much of the observed range.  However, a number of works have found that the CLF is {\em steeper} than the ICMF (e.g., Larsen~2002).  Gieles et al.~(2006a,b) showed that if the the ICMF has a truncation at the high mass end, this will manifest itself as a break (change of index) in the CLF, with the distribution becoming steeper at the high luminosity end.  Such a steepening has been seen in a number of works (e.g., Gieles et al.~2006b; Santiago-Cortes et al.~2010; Bastian et al.~2012; Konstantopoulos et al.~2013; Whitmore et al.~2014).

An additional expectation if the ICMF has a truncation at the high mass end and the star formation is constant over hundreds of Myr, is that the median age of clusters will vary as a function of luminosity, with the brightest clusters being preferentially younger than fainter clusters. This trend is expected because, statistically, the galaxy forms the most massive clusters close to the \mstar, therefore they will have similar masses but their luminosity will fade because of stellar evolution.  For a pure power-law, on the other hand, one would expect that the median age of a sample of clusters is independent of the luminosity.  Larsen~(2009) and Gieles~(2010) exploited this fact and found that brighter clusters were preferentially younger than older clusters, in agreement with expectations if the ICMF is truncated at the high mass end.

\begin{svgraybox}
The ICMF can be approximated by a power-law distribution with an index $-2$ and is stochastically sampled. For many cluster populations the upper end of the mass distribution is better described by an exponential decrease above some characteristic mass, \mstar. Observational evidence and theoretical models suggest that the galactic environment can affect the upper mass end of the ICMF. The chances that the galaxy may form a cluster more massive than \mstar\ are low but not null. Spiral and dwarf galaxies have $\mstar \sim 10^5$~\msun\ while this value increases significantly for cluster populations within galactic mergers and starbursts.
\end{svgraybox}

\subsubsection{The Size-of-sample effect}

\begin{figure}[h]
\includegraphics[scale=0.32]{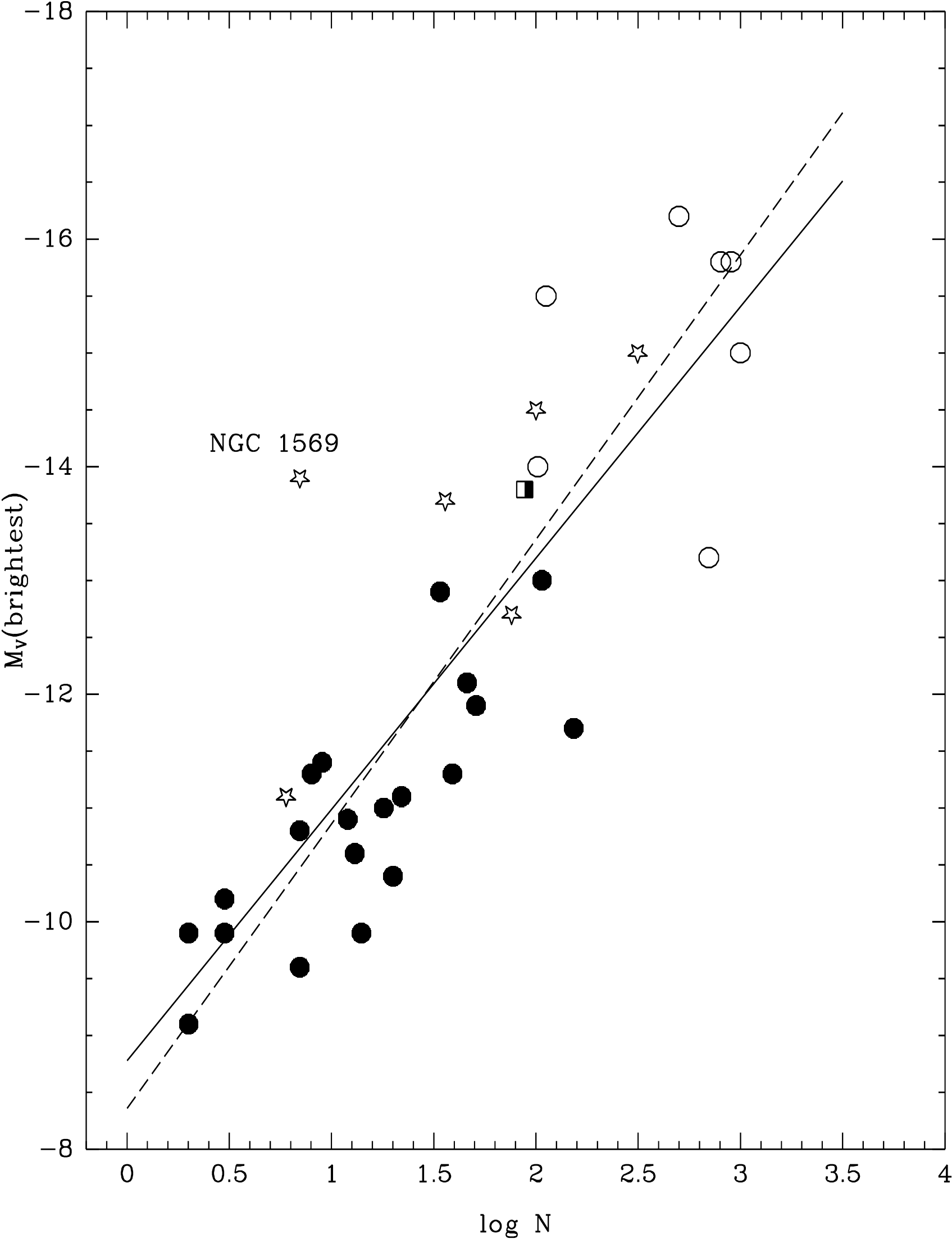}
\includegraphics[scale=0.32]{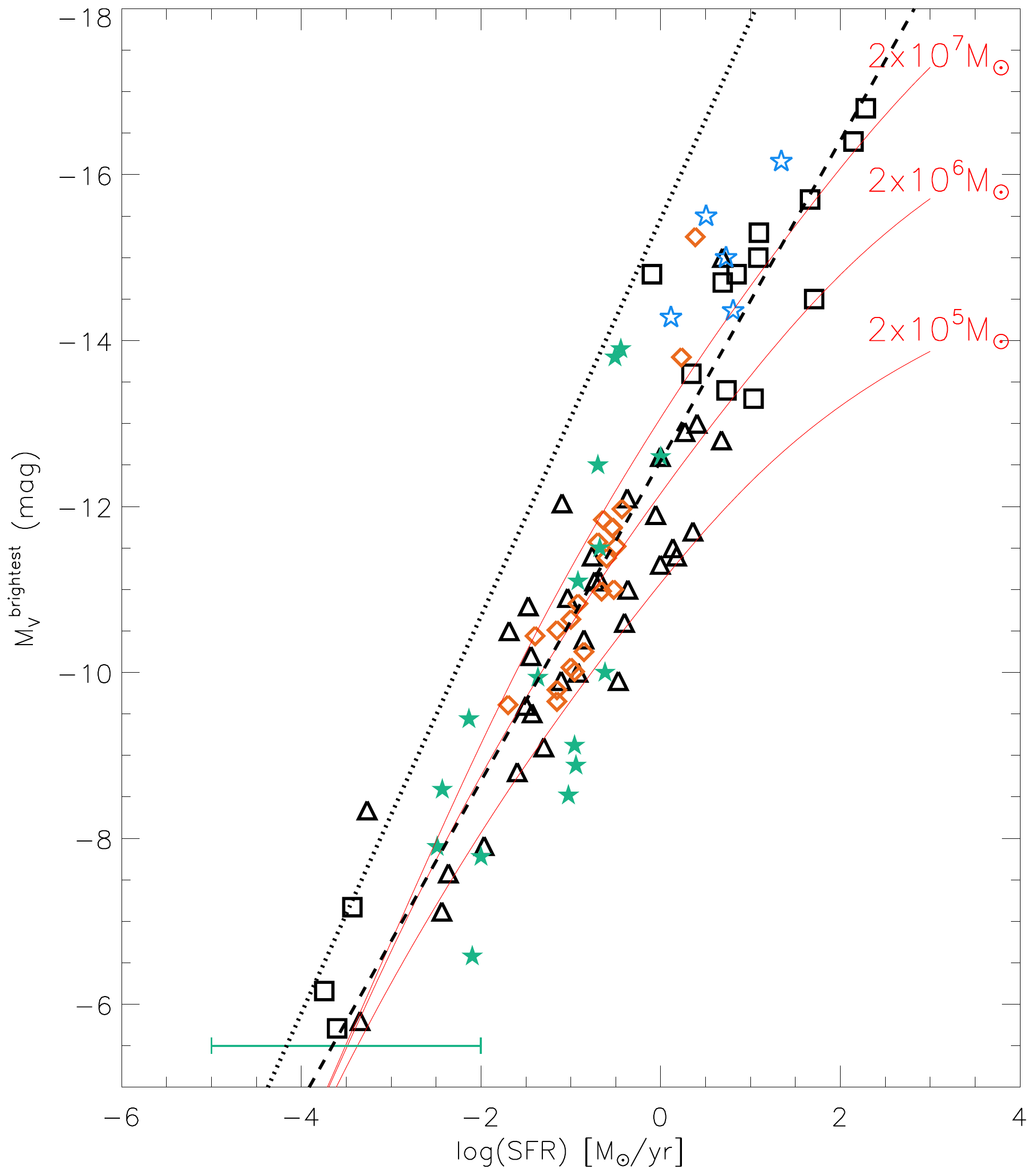}
\caption{{\bf Left panel}: Number of clusters detected in each galaxy versus the $V$ band absolute luminosity of the brightest cluster within the galaxy. The dashed line is the best fit to the data, excluding NGC\,1569, while the solid line is the expected relation if cluster luminosity distribution is described by a power-law, with an index of $-2$ (from Whitmore~2000). {\bf Right panel}: The luminosity of the brightest cluster plotted against the star-formation rate (SFR) of the host galaxy. This plot contains a compilation of all data available in the literature. The dashed line is the best fit to the sample of galaxies plotted as triangles (Larsen~2002). Squares are the sample added by Bastian~(2008).  Blue stars are the sample of luminous blue compact galaxies studied by Adamo et al.~(2011). The orange diamonds show the sample contributed by Whitmore et al.~(2014).    The green stars and the horizontal line represent dwarf galaxies using data compiled from the literature (see text for a detailed description of the data).  Additionally, we show the expected relation for an underlying power-law ICMF ($\alpha=-2$) as a dotted line (for $\Gamma=1$, i.e., 100\% of stars form in clusters) and three relations showing Schechter distributions for the ICMF with three different \mstar\ values (assuming $\Gamma=0.1$).  The plot is taken from Adamo et al. (2015).}
\label{fig:mv_sfr}       
\end{figure}

With the advent of the Hubble Space telescope it was possible to study YSCs not only in the nearby Magellanic Clouds but also in more distant galaxies, probing a much larger range of environments and star-formation rates (SFRs). As the number of samples increased it became evident that the formation of massive star clusters was not only confined to the early universe (i.e. the globular clusters) but that the majority of local star-forming galaxies host YSC populations (similar to some extent to the globular clusters but much younger and less dynamically evolved). Whitmore~(2000) showed that the $V$-band luminosity of the brightest cluster in a galaxy scales with the number of YSCs in the galaxy (left panel, Fig.~\ref{fig:mv_sfr}). He also suggested that the relation could be explained if the clusters were sampled from power-law luminosity (mass) distribution with index $\sim-2$. The nature of this scaling relation became clearer when Larsen~(2002) linked the luminosity of the brightest cluster observed in the galaxy (and the total number of clusters within the population) with the present SFR of the system (see right panel in Fig.~\ref{fig:mv_sfr}). 

Although the formation of clusters must be governed by clear physical processes, with the final cluster properties (e.g., mass and radius) set by the initial conditions and subsequent evolution, cluster populations appear to be stochastically sampled from an underlying parent distribution, the ICMF.  Hence, for higher SFRs, more clusters (i.e. larger populations) are formed.  A more numerous population has a higher probability to sample the cluster mass (luminosity) function at the high mass (brighter) end.  This property of the cluster population is referred to as a size-of-sample effect in the literature, and is the underlying driver of the observed  $M_V^{brightest}$ vs SFR relation.  However, we note that such an effect only dominates a population where the ICMF is not sampled far above the characteristic (Schechter) mass.  In fact, the $M_V^{brightest}$ vs SFR relation implies a steeper ICMF than often found in cluster studies (i.e. an index of $-2.3-2.5$ rather than $-2$ - e.g., Whitmore~2000), implying that a truncation is beginning to affect the relation. {\em While it may appear on face value that this relation may undermine evidence of a truncation or break in the mass/luminosity distributions (\S~\ref{sec:icmf}), the two are consistent given that many galaxies do not sample the ICMF up to the (if present) truncation mass.} In Fig.~\ref{fig:mv_sfr} we show the expected relation between $M_{V}^{\rm brightest}$ and the SFR if the underlying mass distribution is described as a Schechter function with three different values of \mstar\ ($2\times10^5$, $2\times10^6$, $2\times10^7$~\msun).  The implication is that  \mstar\ is related to the SFR, which is expected from theory ( e.g., Kruijssen~2014).  We refer the reader to Larsen~(2010) for a more in-depth discussion of this topic.


The scatter in the $M_V^{brightest}$ vs SFR relation can be understood as being due to the errors associated with the measurements along with the stochastic sampling of the underlying ICMF (e.g., Bastian~2008, see also da Silva et al.~2014).  
For most galaxies the SFR was estimated through its H$\alpha$ flux, which is a measure of the current ($<8$~Myr) SFR of the host galaxy.  In some post-starburst galaxies, however, the current SFR is not a good representation of the star-forming event that formed the highest mass or most luminous clusters.  In extreme starbursts, the most massive cluster formed can be the brightest cluster in the galaxy for hundreds of Myr, especially if the SFR has a sharp decline like in post merger stages or in dwarf galaxies. Therefore, the use of the brightest young (i.e. a cluster that is directly related to the measured star-formation rate) cluster will reduce the scatter. 

It is interesting to note that if one assumes that all stars form in clusters (i.e., 100\% cluster formation efficiency) then we would have expected the observed populations to follow the dotted line in the right panel of Fig.~\ref{fig:mv_sfr}. Using simulated cluster populations by stochastically sampling a Schechter ICMF, Bastian (2008) showed that the observed M$_V^{brightest}$-SFR relation can be reproduced only if a small fraction of the star formation is happening in bound clusters ($\sim8\pm3$\% - see also Gieles 2010). Adamo et al.~(2011), following these results, discussed the possibility that the scatter at high SFR could also be caused by a varying cluster formation efficiency in different galaxies. It is also worth mentioning that many of the highest SFR galaxies either lie at distances where crowding effects may affect the luminosity of single clusters or are in highly extinguished systems (i.e. luminous IR galaxies). On the other hand, if we look at the lowest SFR regimes the scatter is similar, implying that such biases do not strongly influence the results (see Randriamanakoto et al.~2013 for a further discussion).

The $M_V^{brightest}$ vs SFR plot contains the values of about  about 60 dwarfs (total $B$ band luminosity fainter than -18 mag, green stars and green horizontal bar ) which have been searched for clusters\footnote{The green filled stars are a collection of cluster studies in dwarf galaxies of galactic $B$ band luminosity $M_B > -18$ mag \cite{1999ApJ...527..154K, 2002AJ....123.1454B, 2005AJ....129.2701R, 2009AJ....138..169A, 2010MNRAS.405..857G, 2010ApJ...724..296P, 2011AJ....142..129A, 2012ApJ...751..100C, 2013MNRAS.431.2917D}. This sample also contains two systems which have been omitted from the Larsen (2002) catalogue, i.e. NGC\,1569 and NGC1705. These two dwarf starbursts are now included with revised measurement of the galactic SFR \cite[respectively]{2011AJ....141..132P, 2009AJ....138..169A}. } (Fig.~\ref{fig:mv_sfr}). Of this sample about 50 \% of the dwarfs do not have compact bound clusters above the detection limits and about 40\% have clusters (according to the definition of Cook et al~2012) younger than 100 Myr. The galaxies with available cluster photometry and SFRs have been included in the Fig.~\ref{fig:mv_sfr}. A green horizontal bar at the bottom of the plot shows the range in SFR of the dwarf galaxies which do not have bound clusters  (Cook et al.~2012). Some galaxies with similar SFRs have formed clusters where others have not.  It is still under debate if this is an effect of the galactic environment where star formation is happening or whether it is just an effect of the stochastic process at very low SFR regimes (e.g., Cook et al.~2012).

\begin{svgraybox}
The M$_V^{brightest}$-SFR (or number of clusters in a population) relation shows one of the characteristics of cluster populations, that they are dominated by size-of-sample effects. In higher star formation rate regimes, galaxies form more numerous cluster populations which increases the probability to sample the cluster mass (luminosity) function to higher masses (brightness).
\end{svgraybox}

\subsubsection{The Cluster formation efficiency on global scales}
\label{cfe} 
In this section we discuss a relevant aspect of cluster formation and its link to the star formation process. As mentioned in the Introduction, there do not appear to be distinct ``clustered" and "distributed" modes of star formation. Star formation is a clustered process, hierarchical in space and time. Clusters are part of this continuous process and stand out because of their relaxation-dominated dynamics (gravitationally bound structures) emerging at the density peaks within the hierarchy of star formation -- not because of preexisting cloud boundaries (Elmegreen~2006). Massive YSCs usually host a large population of very massive stars, therefore ionising radiation and feedback from clusters may have important effects on galactic scales\footnote{It is currently unclear whether the efficiency of feedback from massive stars is higher if the stars are part of a cluster, rather than being relatively isolated (i.e. in an association) and acting largely on their own.}. To quantify the impact that clusters have on their parent galaxies and at which rate they are formed, it is necessary to probe which fraction of the total stellar mass produced during a star formation event is found in bound YSCs and whether this fraction varies between different galaxies and environments. 

\begin{figure}[h]
\includegraphics[scale=0.3]{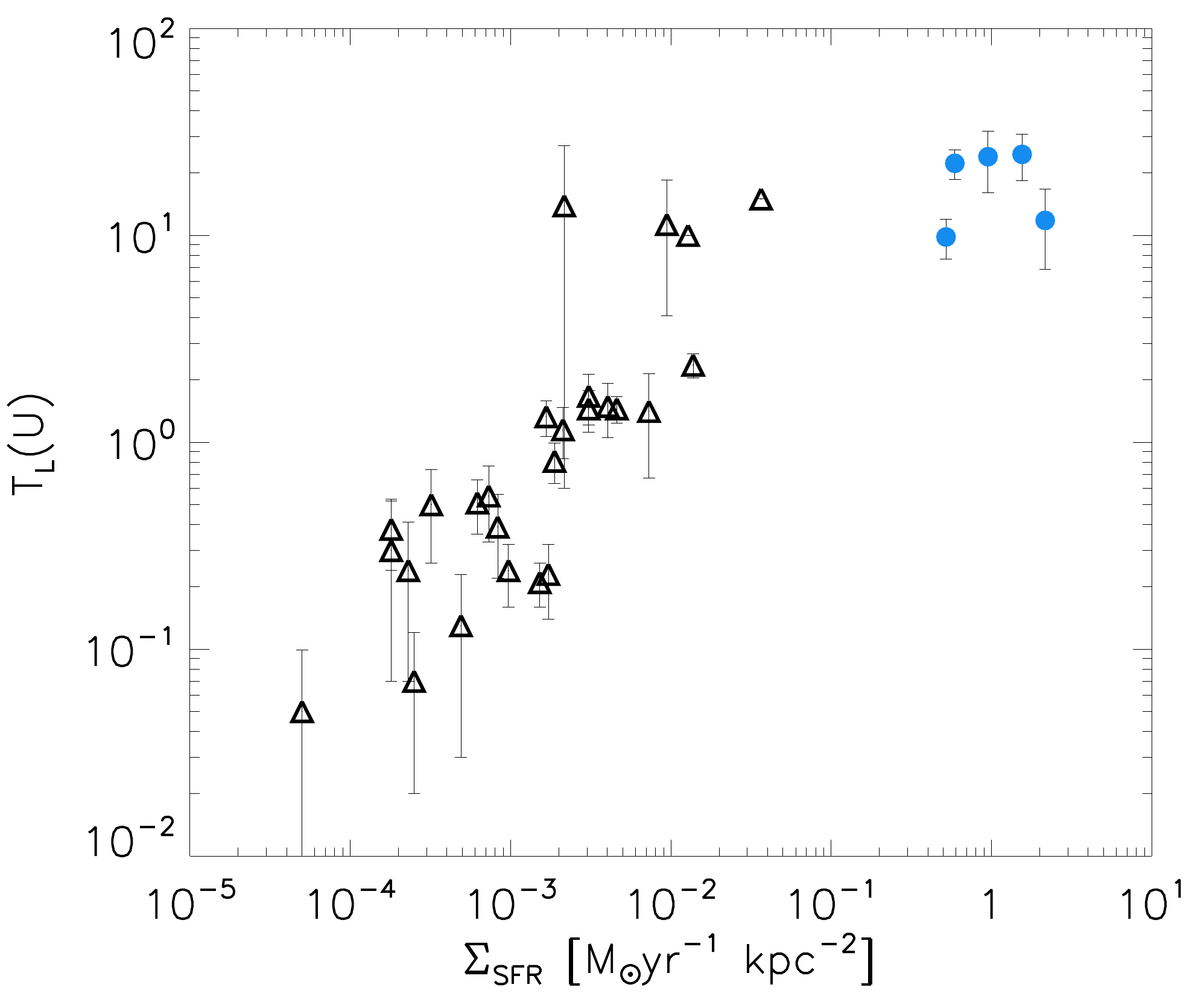}
\includegraphics[scale=0.3]{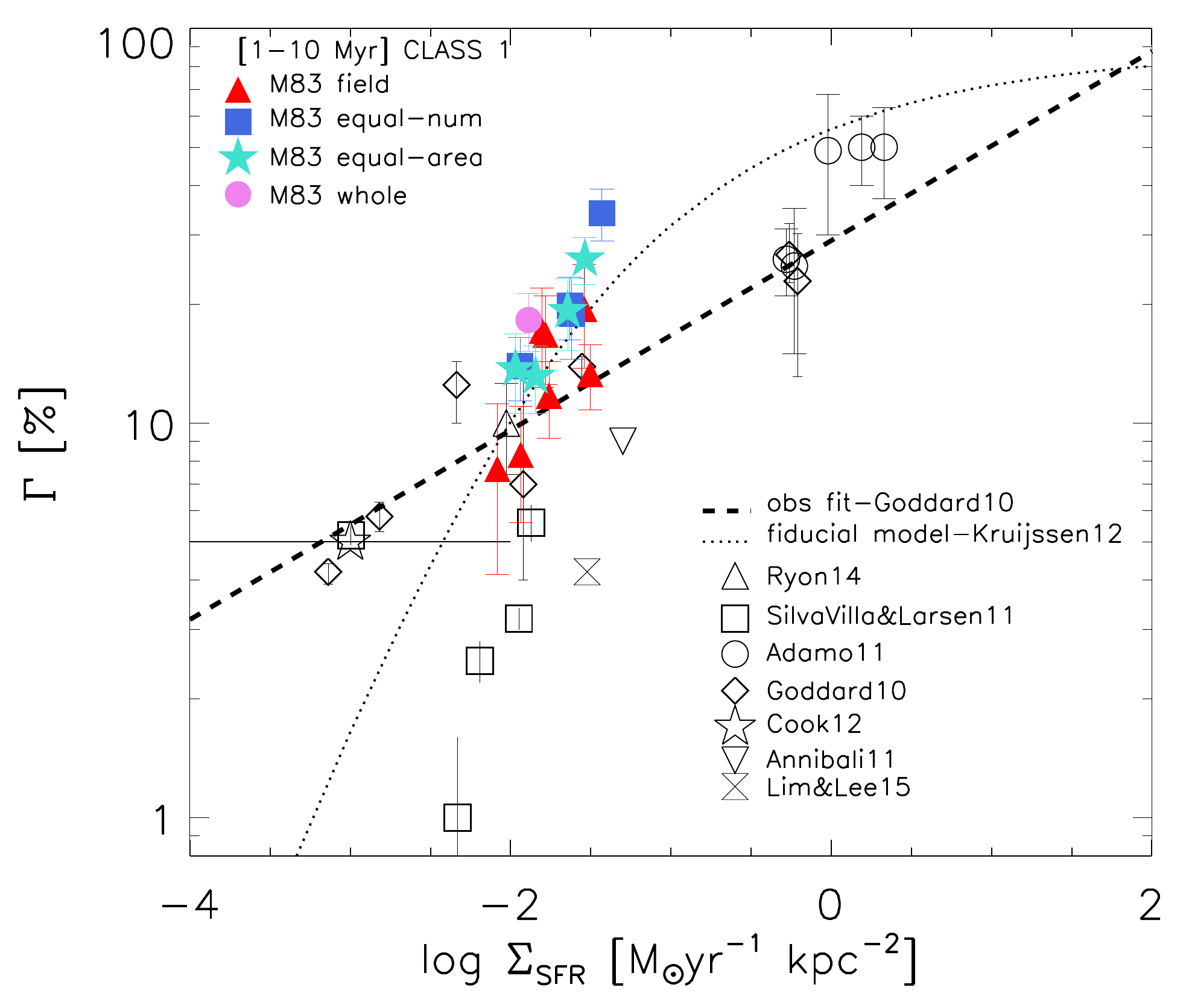}\\
\caption{{\bf Left plot}: The fraction of $U$ band light contributed by YSCs to the total $U$ band luminosity of the galaxy versus the star formation rate density of the host galaxy. Original data from Larsen 
\& Richtler (2000) are plotted as black triangles. The blue solid dots (data from Adamo et al.~2011) extend the relation to much higher SFR density regimes (plot readapted from Adamo et al.~2011). {\bf Right plot}: Cluster formation efficiency ($\Gamma$) versus star formation rate densities. The original plot and datatset by Goddard et al.~(2010) has been updated with all the data available in the literature (see inset). The dashed line is a fit to the Goddard et al.'s data while the dotted line is a fiducial model provided by Kruijssen~(2012). Filled dots are data from a recent study of the cluster formation efficiency in M\,83 on sub-galactic scales. This plot will appear in Adamo et al (submitted). See text for more information.}
\label{fig:gamma}       
\end{figure}

Some of the first ultraviolet (UV) high-spatial resolution images of starburst galaxies provided by HST showed that YSCs dominate the morphological appearance at these wavelengths and significantly contribute ($>20$ \%) to the total UV flux of the galaxy (Meurer et al.~1995).  Larsen \& Richtler (2000) developed a more quantitative approach to the clustering properties of a star-forming galaxy. They used the fraction of luminosity contributed by the YSCs with respect to the total luminosity of the galaxy, in a specific band, i.e. in the UV, T$_L(U)$.  The authors found that T$_L(U)$ increases as function of the averaged SFR density of the host galaxy. In Fig~\ref{fig:gamma} (left panel), we show the original sample by Larsen \& Richtler extended to higher SFR regimes by the luminous blue compact galaxy sample of Adamo et al.~(2011). The scatter in the data is large but the trend is clear. {\em For increasing SFR density, the fraction of stars born in bound clusters is higher. }

The data have not been corrected for any internal reddening (which should not affect the T$_L(U)$ estimates). 
Therefore, there are numerous underlying factors that go into this simple observational relation and their effects have not yet been clearly traced (e.g., the role of a varying SFR). However, the observed increasing trend hints at a tight physical connection between the cluster formation event and the galactic environment where the clusters are forming. 
In the previous section we have discussed the size-of-sample effect. If this process would be the only driving mechanism in cluster formation then we should expect the ratio between the amount of stars formed in clusters and the SFR over the age range of the clusters (this is the meaning of the quantity T$_L(U)$) to be constant. The increasing trend suggests that the cluster formation efficiency (CFE or \gamma) scales positively with the SFR density of the galaxy, or in other words, that the amount of stars born in bound clusters is not a constant fraction but changes as function of the galactic environment. 

A way to probe this statement is to directly look at the cluster formation efficiency in different galaxies. Bastian (2008) define $\Gamma$  as the ratio between the cluster formation rate (CFR) and the SFR. The CFR is usually estimated using the total stellar mass in YSCs over a certain age range. Because of observational limits, the total observed stellar mass in clusters more massive than the limits is used to normalise the ICMF and extrapolate the missing mass hidden below the detection limits, assuming a power-law distribution with index $-2$ (down to $100~\msun$). The SFR is usually derived using indirect tracers like H$\alpha$, FUV and 24 $\mu$m or averaged SFH from direct stellar counts. It is important that the age ranges over which CFR and SFR are estimated are consistent. 

In Fig~\ref{fig:gamma} (right panel), we present a compilation of data available in the literature for which \gamma\ has been measured. The original sample showing the first evidence of an increasing \gamma\, over 5 order of magnitude in SFR densities was originally published by Goddard et al.~(2010). The sample has now been extended to a large variety of galactic environments. The linear fit proposed by Goddard et al. to describe the observed trend (dashed line in the plot) has been replaced by the fiducial model (dotted line) proposed by Kruijssen (2012). The latter model predicts the fraction of star formation that ends up being gravitationally bound by combining different physical processes, i.e. the gas density distribution of the ISM in a galaxy disc, the critical density above which stars form, gas evacuation by star formation and feedback, and the resulting star formation efficiency.  
The flattening at the very high SFR density regimes is produced by the fact that the density of the gas in that regime is so high that nearly only bound structures form.  The \gamma-$\Sigma_{SFR}$ relation, which reflects the more fundamental \gamma-$\Sigma_{gas}$ relation, shows how the galactic environment affects the clustering properties of the star formation process.

\subsubsection{The Cluster formation efficiency on local scales: the case of M\,83}
\label{cfe_m83} 
Silva-Villa et al. (2013) looked for the first time at possible variation of $\Gamma$ within different regions of the same galaxy, M\,83. They find evidence, using the cluster sample from two HST pointings, that \gamma\, declines as a function of galactocentric distances from the centre of the galaxy. 

\begin{figure}[h]
\includegraphics[scale=0.4]{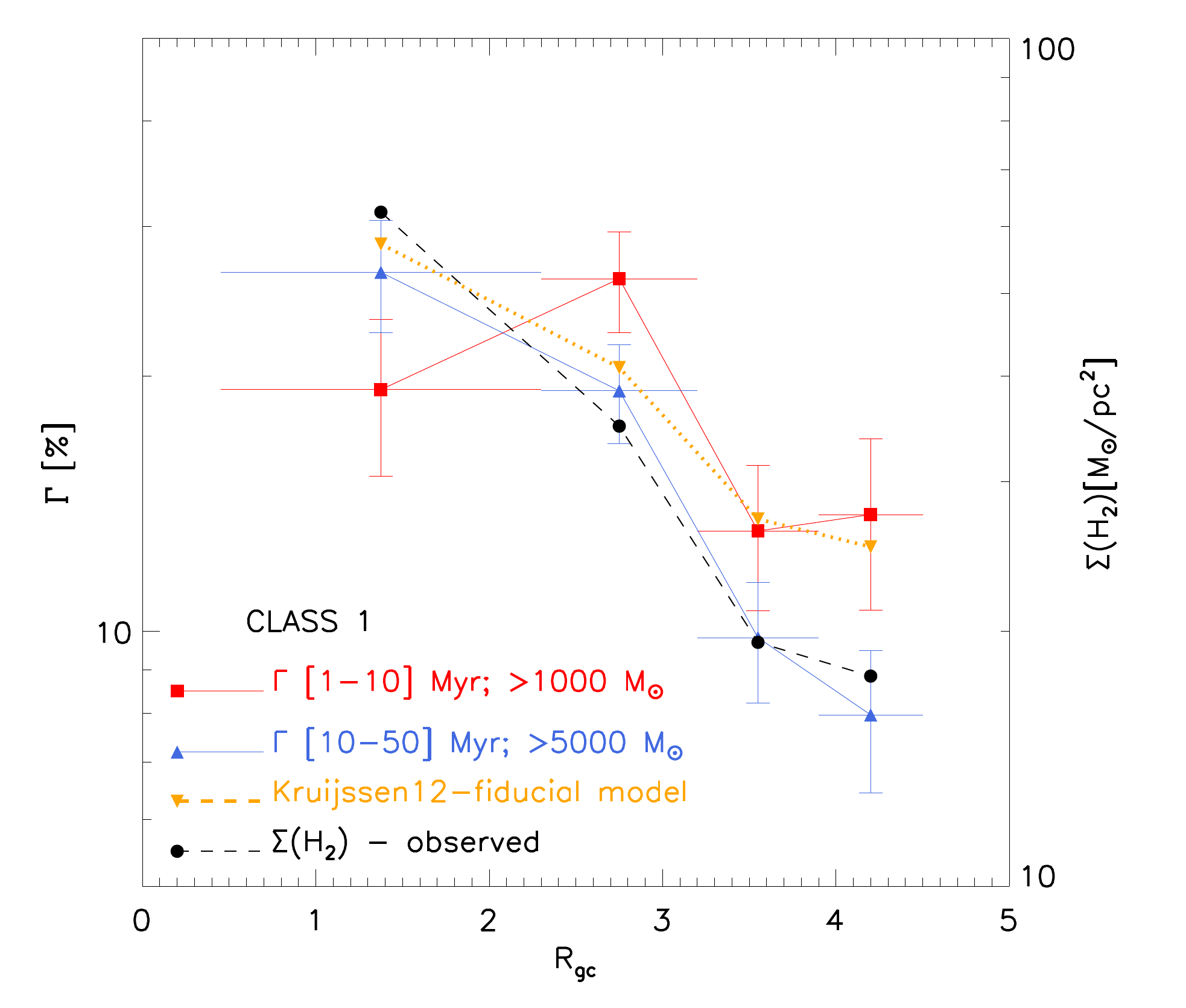}
\caption{The cluster formation efficiency as function of galactocentric distances in the spiral galaxy M\,83. The bins have been selected to have equal area. For clusters younger than $10$~Myr (red solid line with squares) the SFR has been derived using H$\alpha$ as indicator. For clusters in the age range $10-50$~Myr (blue solid line with triangles) the SFR has been derived from resolved stellar populations. Errors on \gamma\, take into account stochastic effects of the ICMF and 0.2 dex in the estimates of the ages and masses of individual clusters. Horizontal bars show the width of the bin. The black dashed line shows the azimuthally averaged gas surface density (right y-axis) in each bin. Taken from Adamo et al. (2015).}
\label{fig:m83_gamma}       
\end{figure}

This analysis has now been extended to the whole galaxy thanks to a complete survey of the M\,83 disk with the exquisite resolution power of the HST (Silva-Villa et al 2014). In Fig.~\ref{fig:m83_gamma} we show how \gamma\, declines as a function of distance from the centre of the galaxy (Adamo et al. 2015). $\Gamma$ has been estimated within annuli of the same area. The detection limits used to estimate the observed total stellar mass in clusters (the amount in clusters less massive than this limit is inferred assuming a power-law ICMF) is a function of the age range considered. The SFR compared to clusters younger than 10~Myr has been estimated from H$\alpha$ images, while the SFR for clusters with ages between 10 and 50~Myr is derived from direct stellar counts. 
Note the systematic decrease in $\Gamma$ as a function of galactocentric distance.  To reinforce the link with the underlying galactic environment we overplot the azimuthally averaged gas surface density measured in each annulus. The correspondence between the radial variation of $\Gamma$ and gas surface density profiles was quantitatively predicted by the model of Kruijssen~(2012, yellow triangles in Fig.~\ref{fig:m83_gamma}), where the fraction of stars bound in clusters is a function of the molecular gas surface density, which is near-linear at low ($\Sigma\leq 50~\msun$) surface densities.
{\em Hence, we conclude that the fraction of stars that are formed in bound clusters depends on the local and global environment, and ranges from $\sim3$\% (or less) in quiescent dwarf galaxies up to $\sim50$\% or more in intense starbursts.}


\begin{svgraybox}
Observations and theoretical models have found that the clustering properties of the stellar population change as a function of the galactic environments. Higher SFR densities produce on average larger $\Gamma$, i.e. a larger fraction of the star formation is happening in bound clusters.
\end{svgraybox}

\section{The Cluster Age Distribution and Cluster Disruption}

Early work with HST led to the exciting conclusion that major starburst events within galaxies result in the formation of hundreds of massive, globular cluster progenitors (e.g., Holtzman et al.~1992; Miller et al.~1997).  Hence the age distribution of clusters held extraordinary potential to derive the star-formation history of galaxies, or at least their major star-forming episodes.  However, it was known that clusters do not survive forever, but rather lose mass through a variety of processes, discussed in more detail below (e.g., Spitzer~1987).  Correcting for this cluster mass loss or disruption has become a point of major contention in the field.  Below, we outline the basic physical properties, expectations, debate on the empirically derived disruption laws, and summarise the current observational state of the field.

\subsection{Expectations from Theory and Parameterisations}

Once a cluster forms, a number of processes cause the cluster to lose mass (i.e., lose stars from the cluster to the surroundings), eventually leading to its entire disintegration.  If the cluster forms, and the gas left over from the non-100\% star-formation efficiency makes up a significant amount of the mass of the cluster (i.e. the gravitational potential is still dominated by the gas), then the removal of this gas, on a short timescale, may cause the cluster to lose much of its stellar mass, potentially disrupting the entire cluster (e.g., Lada et al.~1984), in a process known as ``infant mortality".  Recent observations (e.g.) as well as numerical simulations (e.g., Kruijssen et al.~2012) suggest, however, that massive clusters are not strongly affected by this process (see Longmore et al.~2014 for a full review), so we shall not deal with this process in detail here.  However, we note that for massive clusters, even if gas expulsion does modify the cluster, the cluster will be back in equilibrium within 5-20 Myr (Longmore et al.~2014).  

A potentially much more severe disruption process is caused by the interaction of young clusters with GMCs in their vicinity.  Since clusters are born in gas rich environments, this effect will be strongest at young ages and will decrease as the cluster moves away from its natal star-forming region (Elmegreen 2010; Kruijssen et al.~2011).  This process is often referred to as the ``cruel cradle effect".  If the density of GMCs is high, the gravitational shocks imparted by the GMCs on the young clusters are expected to be strong. Sufficiently strong shocks could disrupt any cluster in a single encounter, leading to mass-independent cluster disruption.  Under less extreme conditions, the mass loss on the cluster is expected to be proportional to the cluster density, with lower density clusters easier to destroy.  Since, YSCs do not, in general, display a mass-radius relation (e.g., Larsen 2004), this means that this process should be proportional to mass, so higher mass clusters should live longer.

If a cluster survives long enough to escape from its natal gas-rich environment, it will still lose mass due to 1) the gravitational tidal field of its host galaxy, 2) encounters with GMCs, 3) stellar evolution and 4) two-body relaxation (an internal process - although governed by the external tidal field - Gieles \& Baumgardt~2008).  The relative strength of the first two processes depends on the environment.  It is beyond the scope of this chapter to discuss these processes in detail, and we refer the interested reader to the excellent review by Portegies Zwart et al.~(2010) as well as the detailed discussions provided by Lamers et al.~(2010) on the tidal field and Kruijssen et al.~(2011) on tidal shocks (c.f., their Fig.~8).  In principle, one can tune the above processes to make them all (nearly) independent of mass (e.g., Fall et al.~2009), however, the first two will always remain strongly environmentally dependent.  {\em The basic outcome of theory is that in most environments, more massive clusters should survive for longer and that in environments with high GMC density and/or strong tidal fields cluster dissolution should happen more rapidly (for a given cluster mass).}

\subsection{Analysing Cluster Populations}

Throughout this section we will only discuss {\em mass-limited samples}.  It is possible to use {\em luminosity-limited samples}, e.g., Boutloukos \& Lamers~(2003), however it complicates the analysis.  Many apparent contradictions in the field can be traced to the use of luminosity-limited samples being analysed as if they were mass limited.  We will discuss the behaviour of luminosity-limited samples when necessary.

We will, following on from previous works, approximate the cluster age distributions as power-laws, normalised to the linear range of the age bin, namely of the form $\dndt \sim t^{-\zeta}$.  In this form, if the cluster formation rate is constant and no disruption acts on the population, then the distribution should be flat (i.e. constant) with age, $\zeta=0$.  If disruption affects a population, then the distribution should become steeper at older ages, as young clusters have not undergone much mass-loss relative to older clusters. However, if a sample is luminosity limited, this also steepens the age distribution, and can lead to erroneous conclusions regarding the role of cluster disruption in shaping the observed population\footnote{If a sample is luminosity limited, the age distribution for the case of no disruption and a constant cluster formation rate is expected to decrease with $\zeta=0.65, 0.9$ if the sample is limited in the V or U-bands, respectively \cite{2010ASPC..423..123G}.}.

In the literature, two empirical disruption laws have been advocated, {\em mass independent disruption} (MID - e.g., Whitmore et al.~2007) and {\em mass dependent disruption} (MDD - e.g., Lamers et al.~2005).  As their names suggest, the two scenarios propose different dependencies of the cluster mass on the cluster lifetime. They also predict different roles of the galactic environment. I.e., the MID scenario assumes that cluster disruption has little or no dependence on environment, while the MDD predictes a strong dependence on environment.  We refer the reader to the review contained in Bastian et al.~(2012) for a more thorough discussion of the models. 

These two disruption laws, produce clear differences in the expected age distribution (see Lamers~(2009) for an in depth discussion).  Briefly, the MID scenario predicts that because cluster disruption is independent of cluster mass and the local environment, all age distributions should be similar (modulo SFH effects), following a single power-law with index, $\zeta \sim 0.9$ (e.g., Whitmore et al.~2007).  At what age this rapid decline should stop, is still an open question.   For the MDD scenario, due to the dependence of cluster disruption on the local environment, we would expect to see a range of age distributions, additionally we should see not a single power-law, but rather multiple parts to the distribution.  At young ages, when disruption has not acted strongly yet (modulo ``infant mortality" and the ``cruel cradle effect") the age distribution should be flat ($\zeta \sim 0$).  This should then steepen at older ages, as cluster disruption begins eating into the population.

\begin{svgraybox}
Theoretically, cluster disruption is quite well understood, with the rate of cluster disruption, for a given mass, dependent on the ambient environment.  If the tidal fields are strong or large numbers of GMCs are present, the lifetimes of clusters should be significantly shorter than in environments with weak tidal fields or few GMCs.  In the case of strong disruption, the age distribution of clusters should be steeper than in the case of little or no disruption.
\end{svgraybox}

\subsection{Numerical results}


Kruijssen et al.~(2011; 2012, hereafter K12) ran a series of galaxy scale gravitational and hydrodynamical models of quiescent and merging spiral galaxies.  In these simulations, clusters were allowed to form from the gas if the local density exceeded some threshold density.  The gas in this region was then converted to stars in clusters, and the clusters were sampled from a power-law mass function with index, $-2$.  The evolution of these clusters were then followed in a sub-grid model, taking into account their  galactic environment and the dissolution effects discussed above.  All of their cluster mass-loss algorithms were calibrated to direct N-body simulations of clusters with stellar evolution in a tidal field.  

\begin{figure}[t]
\sidecaption[t]
\includegraphics[scale=0.4]{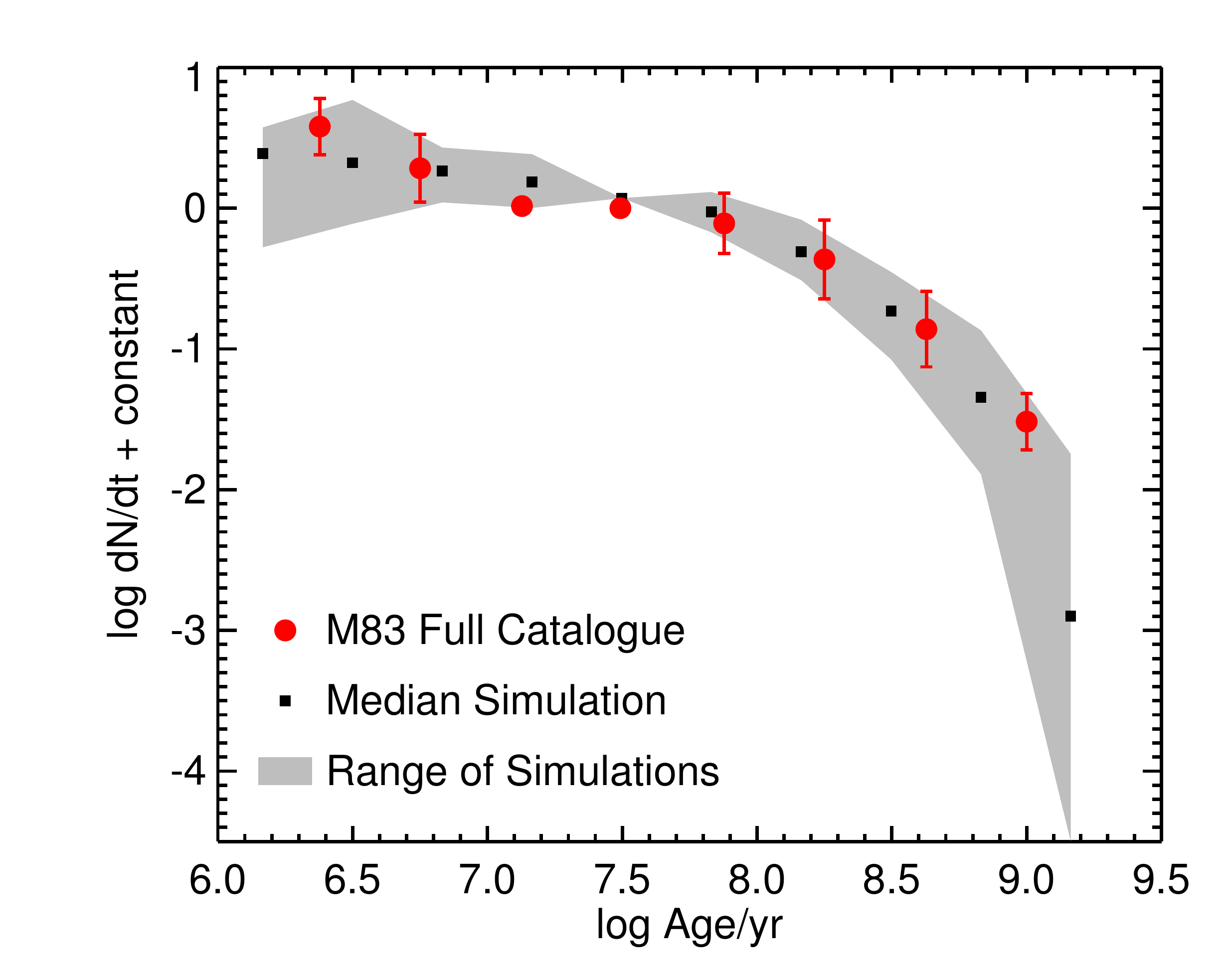}
%
%
\caption{A comparison of age distributions from the numerical simulations (solid squares represent the median simulations and the shaded region shows the full range of the simulations) of cluster populations in spiral galaxies (Kruijssen et al.~2012) and the observed cluster population of M83 (Silva-Villa et al.~2014).  Both distributions are normalised at 30~Myr, to allow a direct comparison.  We show the full cluster catalogue (applying a lower mass limit of 5000~\msun) and the error bars represent the differences between two similar fields (F1 and F5) were the disruption timescale should be comparable.  The up-turn in the observational data at young ages ($<10$~Myr) may be due to some amount of cluster disruption (i.e., infant mortality/cruel cradle effect) or the inclusion of associations in the catalogue. }
\label{fig:numerical}       
\end{figure}

The authors found that in gas-rich mergers, cluster disruption could indeed proceed largely independent of the cluster mass.  However, in their quiescent spirals, cluster disruption was a much slower process, showing a clear environmental dependence, along with a dependence on the cluster mass.  In Fig.~\ref{fig:numerical} we show the median age distribution of the cluster population of twelve of the K12 quiescent spiral galaxies as filled black squares, and the grey shaded region shows the full distribution found in the models.

Each of the galaxies in the K12 simulation shows the same overall trend.  A near-flat part of the distribution (i.e. where disruption is not strongly affecting the population) and then a downwards curve.  Since we have a mass limited and complete sample, this effect is entirely due to disruption.  Environments where disruption is faster will have age distributions that bend earlier, compared to environments where disruption proceeds slower.  Unfortunately, incompleteness also can cause the downward bend at old ages, so care must be taken when analysing observed age distributions.

Hence, the results from these numerical simulations agree with the expectations from analytical theory (e.g., Lamers et al.~2005; Lamers \& Gieles 2006).  The models of Renaud \& Gieles~(2013) largely confirm the results of the Kruijssen et al. (2011; 2012) simulations for the gas-poor part of the parameter space where they overlap -- as the former
have not included gas (GMCs) in their simulations so they find weaker cluster disruption than in the gas-rich environments included by the latter simulations.  While the theory behind cluster disruption appears to be on strong footing with little debate, the observational picture is more complicated, and has been the subject of an ongoing debate within the literature (MID vs. MDD).  Below we discuss the observed age distributions of clusters in different environments, and compare studies done by different groups.

\begin{svgraybox}
Numerical simulations and analytic theory predict that the age distribution for quiescent spirals should show a flat portion, from young ages to $\sim100-300$~Myr, followed by a steeper portion where cluster disruption is dominating the population.
\end{svgraybox}

\subsection{Observational Results on the Cluster Age Distribution}

There has been a significant amount of work done on cluster populations in the Galaxy, as well as nearby galaxies, especially since the advent of HST.  However, the past decade has also witnessed a significant amount of controversy regarding this topic, which in turn has strongly impacted the discussion of the lifetimes of clusters.  As discussed above, if the lifetimes of clusters are short (tens of Myr or less), then the overall population age distribution will be steep, at least over the timeframe where disruption is occurring.  If, on the other hand, clusters are stable when they form, and survive for hundreds of Myr, then the age distribution is expected to be shallow.   However, as we will see, a single power-law is not a good description of many of the cluster populations studied to date, so we will be paying particular attention to the age range over which the fit was carried out.

It is also important to remember that the overall SFH of the galaxy can influence the age distribution of the clusters (see, e.g., Bastian et al.~2009).  If the SFR of a galaxy has been increasing the age distribution will become steeper, whereas it will become flatter (or even inverted) if the SFR has been decreasing.  Clearly, the assumption of a constant SFR for merging or starburst galaxies is questionable, whereas this should be a better assumption when looking at the full cluster population in more quiescent spirals.


In this section we look at a number of results from the literature, and study a handful of cluster systems in detail as case studies.

\subsubsection{The Open Cluster Population in the Milky Way}

Our knowledge of the open cluster population of the Galaxy is surprisingly incomplete.  Piskunov et al.~(2006) suggest that we are only complete out to a distance of $\sim800$~pc from the sun.  This limit is important, as samples of clusters out to, e.g., $\sim2$~kpc are incomplete, and behave as luminosity-limited samples.  An example of such a behaviour can be seen when comparing the age distribution of open clusters in Lada \& Lada~(2003 - based on the catalogue of Battinelli \& Capuzzo-Dolcetta~1991) with that of Lamers et al.~(2005) or Piskunov et al.~(2006).  Lada \& Lada~(2003) find that the number of clusters per {\em logarithmic} bin is roughly constant with age, which suggests that the $\dndt \propto t^{-1}$, i.e., $\zeta=1.0$.  The authors conclude that up until $\sim100$~Myr, 90\% of clusters disrupt every decade of age, i.e. very strong cluster disruption.  However, the catalogue used included clusters out to $2$~kpc, hence was effectively luminosity limited.

In comparison, Lamers et al.~(2005, also see Piskunov et al.~2006) found that the age distribution was largely flat to an age of $\sim100$~Myr and then rapidly decreased, if a mass-limited sample was used, including only clusters within $800$~pc of the Sun.  The authors used the MDD framework discussed above to conclude that a cluster with a mass of $10^4$\msun, will (on average) survive for 1.7~Gyr in the solar neighbourhood.  Hence, it appears that in the solar neighbourhood, stellar clusters are long lived entities, in agreement with expectations given the relatively weak tidal field and the scarcity of massive and dense GMCs.

\subsubsection{The Cluster Population of M31}

A recent survey that deserves special consideration is the {\em Panchromatic Hubble Andromeda Treasury} (PHAT) survey, which covers a 0.5~deg$^2$ area of M31, extending from the central regions out $\sim20$~kpc (Dalcanton et al.~2012).  Johnson et al.~(2012) have analysed the ``1st year data" of the survey, which covers five ``bricks" (collections of HST imaging footprints) from the inner to the outermost regions of the galaxy, and presented integrated luminosities in six filters for 601 clusters identified in their sample.  Fouesneau et al.~(2014) used this sample to estimate the ages, masses and extinctions of the clusters using stochastic SSP models and a Bayesian analysis method.  They then construct age distributions for three radial bins at 6, 10 and 15~kpc, and find a flat distribution ($\zeta \sim 0$) for the first $\sim70-100$~Myr, after which the distribution declines rapidly, with $\zeta=1.15$.  Remarkably, each of the three fields shows the same distribution.  The rapid decrease after $100$~Myr is due to a combination of their completion limit (i.e. the sample becomes luminosity limited after this age) and cluster disruption.  However, it is clear that there is little evidence for rapid cluster disruption within M31 (at these radii) for at least the first $100$~Myr.  As was found for the solar neighbourhood, and in numerical simulations, it appears that once a cluster forms, it is a long lived entity in the Andromeda galaxy.

While the full survey is expected to add an additional $\sim2000$ clusters to the sample, and will place the results on an even stronger statistical footing, it is clear from the current data that the population follows the expected trends, and that rapid cluster disruption within the first 100~Myr is inconsistent with the data.

\subsubsection{The Cluster Population of the LMC}

The LMC is the nearest galaxy to us with a significant young cluster population, and as such has been the subject of numerous studies.  Here we only focus on results from the past $\sim5$ years, given the controversy that has emerged on the issue of the age distribution in this galaxy.  Chandar et al.~(2010a) used the cluster catalogue of Hunter et al.~(2003) and re-estimated each cluster's age, mass and extinction.  The authors find that for ages between $1$ and $1000$~Myr, the age distribution can be well described by a single power-law with $\zeta=0.8$.  Unfortunately, the data used for their analysis has not been made publicly available, so it is not possible to confirm the results.  Chandar et al. also find  a relatively large population of massive ($>10^4$~\msun) young ($<10$~Myr) clusters in their sample, i.e. eight R136 type clusters.  Given the ease of detecting these kinds of objects, and their lack of appearance in other studies, it seems likely that these are misfit clusters, leading to an overestimation of the number of such very young massive clusters in the LMC.



Baumgardt et al.~(2013) collated all major publicly available catalogues of clusters in the LMC, removing a significant amount of double detections (often within the same catalogue) and re-estimated each cluster's age, mass and extinction. The authors only include clusters older than 10~Myr. These authors find a significantly different distribution than that reported in Chandar et al.~(2010a), namely a flat age distribution to ages of $200-300$~Myr ($\zeta \sim 0.3$), followed by a steep decline (again caused by a combination of disruption and incompleteness). de Grijs et al.~(2013) independently collated cluster studies of the LMC, and found results consistent with Baumgardt et al.~(2013) and inconsistent with Chandar et al.~(2010a). 

Comparing the Chandar et al. and Baumgardt et al. distributions, it appears that some difference is caused by the choice of binning, with the youngest age bin of the Chandar et al. study forcing the fit to steeper values, as the age range between $10-100$~Myr is largely flat in their sample.  This highlights the danger of adopting a single value for the binning of data, showing that at least multiple bin widths need to be considered, or, preferably, better statistical analyses such as maximum likelihood comparisons.  We have carried out a maximum likelihood fit on the Baumgardt et al. sample, fitting the age distribution over different age intervals (for mass-limited samples, $M>5000$~\msun). For the age interval from 10-100~Myr, we confirm that Baumgardt et al. value of $\zeta=0.35$.  Once older ages are included, the age distribution begins dropping rapidly (likely due to a combination of disruption and incompleteness).  Fitting the full range from 10-1000~Myr, we find $\zeta=0.9$, in good agreement with Chandar et al.

The obvious interpretation of these results is that a single power-law fit to the data is not a good representation to the cluster population of the LMC.  For ages younger than 100~Myr, there appears to be no evidence for rapid disruption (c.f. Baumgardt et al.~2013; de Grijs et al.~2013).  For older ages, disruption and incompleteness are likely causing the steepening the age distribution.

\subsubsection{The Cluster Population of M83}

Due to its proximity and large amount of HST/WFC3 coverage, the spiral galaxy, M83, has been targeted by a number of recent cluster studies.  An additional importance of this galaxy is that due to its distance ($\sim4.5$~Mpc), it is possible to sample different environments within the same galaxy (with a reasonable amount of observing time), while still semi-resolving the clusters.  Hence, it is an excellent environment to test the environmental dependence of cluster disruption.

\begin{figure}[t]
\includegraphics[scale=0.4]{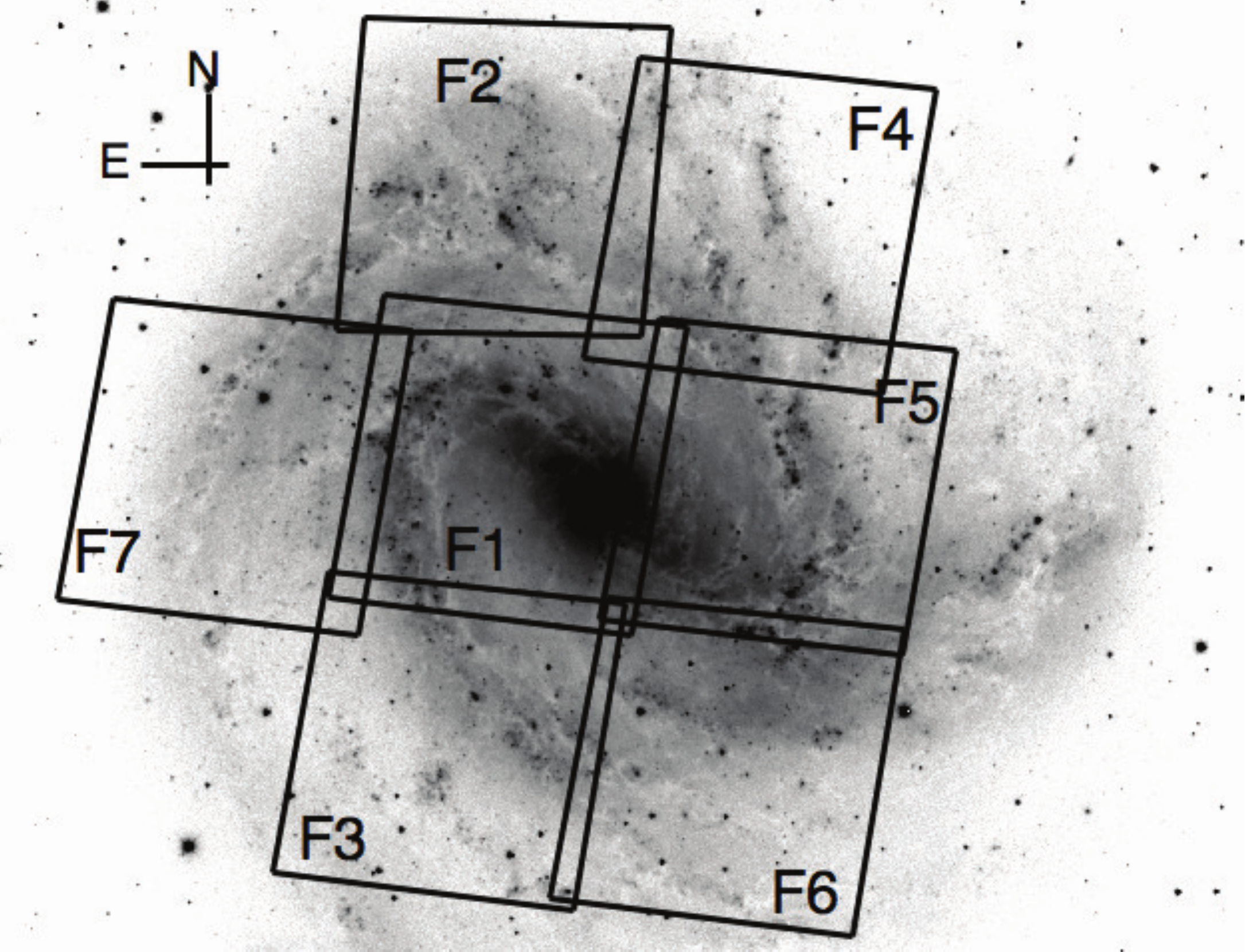}
%
%
\caption{An R-band image of M83 with the seven HST/WFC3 fields superimposed and labelled (taken from Silva-Villa et al.~2014).}
\label{fig:m83_image}       
\end{figure}

Chandar et al.~(2010b) studied the first of seven fields, F1 (see Fig.~\ref{fig:m83_image}) with multi-wavelength HST/WFC3 imaging, covering the inner region of the galaxy.  Using similar methods to those discussed above, they found $\zeta=0.9$ from $1-1000$~Myr, for a single power-law fit.  Bastian et al.~(2012) reanalysed F1, and overall, found excellent agreement with both the cluster catalogue and derived properties, and also the age distribution, finding $\zeta=0.85$ over the same age range.  The main differences between the catalogues were restricted to young objects ($<10$~Myr) as it is difficult to distinguish between bound clusters and unbound associations at these ages (e.g., Gieles \& Portegies Zwart~2011).  Bastian et al. simply adopted more conservative criteria for identifying clusters, although this is largely a subjective distinction.  Hence, the Bastian et al. sample provides lower limits at young ages, while the Chandar et al. sample provides upper limits for the age distribution.  For ages older than $10$~Myr, the two populations gave nearly identical results.

\begin{figure}[t]
\includegraphics[scale=0.4]{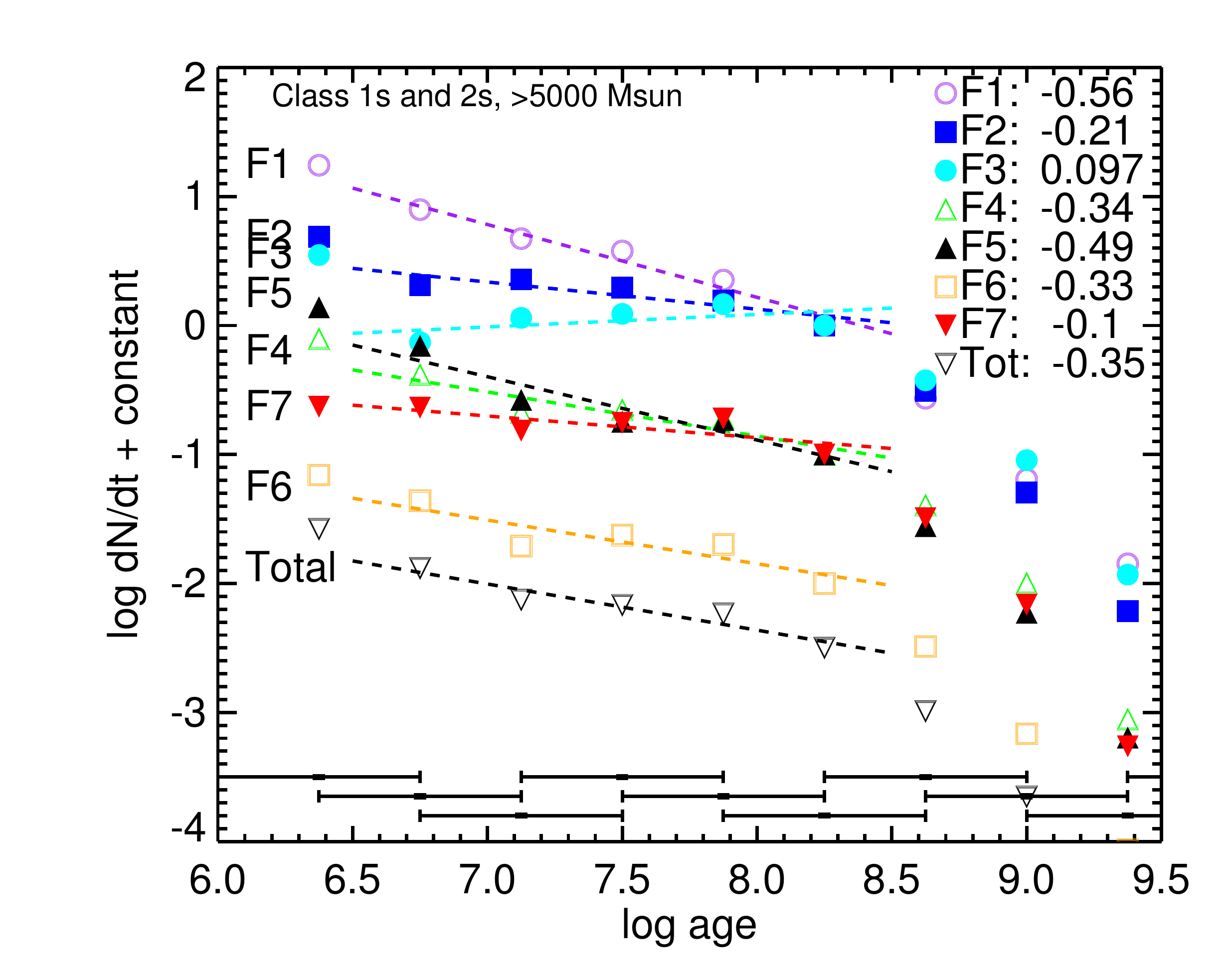}
%
%
\caption{The age distribution of clusters and associations in seven (slightly overlapping) fields (each one HST/WFC3 pointing).  The lines indicate the best fit slope ($-\zeta$) over the range indicated (for mass-limited samples), and the values are listed in the panel.  F1 covers the central part of the galaxy, and has the steepest slope, while the fields that cover the outskirts of the galaxy (e.g., F2 and F7) display significantly shallower slopes.  This clearly shows the environmental dependence of cluster disruption.  The drop at  $\sim200$~Myr is a combination of incompleteness and cluster disruption (taken from Silva-Villa et al.~2014).}
\label{fig:m83_dndt}       
\end{figure}

However, Bastian et al. also studied a second field, F2 (see Fig.~\ref{fig:m83_image}) using the same techniques, and found that the age distribution there was significantly shallower ($\zeta=0.4-0.5$).  This is expected if cluster disruption is environmentally dependent, as further from the galaxy centre, the tidal field and the number of GMCs have dropped considerably, meaning that clusters are likely to survive for longer (e.g., Lamers et al.~2010; Kruijssen et al.~2011).  Bastian et al.~(2011) showed that even in colour-space (i.e. before an age dating is done) the clusters in F2 are significantly redder in $U-B$ than those in F1, showing that they have older ages (extinction can not cause the observed colour differences).

Chandar et al.~(2014) also studied F2, and found results consistent with those of Bastian et al.~(2011; 2012). However, the authors suggest that the differences between the two fields is only at the $2-3\sigma$ level.  The Chandar et al.~(2014) data are public so we can look into this issue in detail.  One difference between the Chandar et al. (2010) and (2014) results was that in 2014, only clusters older than 10~Myr were included in the analysis.  However, if younger clusters are included so that we analyse the age range of $5-300$~Myr (as the two fields were treated equally, the distinction between clusters and associations should not affect the results), the two fields have very different distributions.  The age distribution of F1 is much much steeper than F2, with $\zeta_{F1}=0.85\pm0.15$ and $\zeta_{F2}=0.15\pm0.15$ (in agreement with the independent analysis done by Silva-Villa et al.~2014).  Using a KS-test, we find that the two samples have a probability of $<1 \times10^{-5}$ of being drawn from the same parent distribution.

Hence, it appears that all catalogues of M83 studied to date agree that the cluster population closer to the galaxy centre is different than in the outer parts, in the way predicted by environmental dependent cluster disruption theories.  Results suggesting otherwise were largely caused by the choice of age range over which the fit was carried out.

Finally, Silva-Villa et al.~(2014) have studied all seven fields, using the same techniques, and found that $\zeta$ varies from $0.8$ in the central regions to $\sim0$ in the outer regions.  Their age distributions for the seven fields are shown in Fig.~\ref{fig:m83_dndt}.  The authors found excellent agreement when comparing their results to the Chandar et al. (2010; 2014) and Bastian et al.~(2012) results.  The Silva-Villa et al. full catalogue is shown in Fig.~\ref{fig:numerical} in comparison with the simulations of Kruijssen et al.~2012.  Note the excellent agreement with the simulations, which explicitly predict that cluster disruption is dependent on both the environment and the initial cluster mass.

We conclude that the age distribution in M83 is clearly dependent on location within the galaxy.  The inner regions of the galaxy are characterised by relatively steep age distributions, indicative of heavy disruption.  However, in the outer regions of the galaxy the age distributions are significantly shallower (in some cases, nearly flat).  As discussed in Bastian et al.~(2012) (and above) this is in excellent agreement with predictions of environmentally dependent cluster disruption (MDD).

\subsubsection{Other Cluster Population Studies From the Literature}

While we have focussed on a handful of cluster populations in detail, a number of other studies have found clear evidence that the age distribution of clusters depends systematically on the ambient environment.  Galaxies with strong tidal fields and/or large GMC populations have steeper age distributions, while galaxies, like the SMC, where cluster disruption is not expected to be a strong effect, have flat distributions.  In Table~\ref{tab:1} we show the results of other recent works from the literature as well as for the galaxies discussed in the previous sections.  We also highlight the age range over which the fit was carried out.  This can, as discussed above, strongly affect the resulting fits, as the inclusion of unbound associations at young ages and/or the inclusion of ages older than the completeness limit allows, can lead to significantly steeper distributions than is physically present.  

{\em From this growing list of studies it is clear that the age distribution of clusters is not universal, but rather depends strongly on the ambient environment.  However, care must be taken when fitting the distributions, as approximating the full age distribution by a single power-law over the full observed range can lead to erroneous conclusions.}

\begin{svgraybox}
There has been a significant amount of debate in the literature on the form of the age distribution of cluster populations, which in turn has led to uncertainties in the role of cluster disruption in shaping the population. Publicly available catalogues have been used to compare results between different teams and galaxies, and now clearly show ($P_{KS} < 10^{-5}$) that the age distribution varies strongly as a function of environment, with some galaxies (or regions) having flat ($\zeta \sim0$) distributions  (i.e., little disruption) while others display evidence of steep declines ($\zeta \sim 1$), indicative of strong disruption.  Environments with weak tidal fields and/or low numbers of GMCs show flatter age distributions, consistent with analytical and numerical expectations.
\end{svgraybox}


\begin{table}
\caption{List of measurements of the cluster age distribution in different galaxies, focussing, with the exception of the Antennae galaxies, on systems where the SFH should have been largely constant over the age range measured. Throughout, we have assumed a power-law type profile of the form $\dndt \sim t^{-\zeta}$ over the age range listed.  $^a$Based on the upper envelope of the age-mass relation (see Gieles \& Bastian~2008).  $^b$Similar results have also been found by \cite{2007AJ....134..447S}, \cite{2014ApJS..211...22F}, and \cite{2015A&A...581A.111D}. However \cite{2007AJ....134..447S} used a luminosity-limited sample, hence they erroneously interpreted their steep distribution as being caused by disruption, correcting for this leads to $\zeta\sim0.4$. $^c$Note the difference between this result and \cite{2006ApJ...650L.111C} who effectively used a luminosity-limited sample, hence found a much steeper age distribution.  The Gieles et al. result was independently confirmed by \cite{2008MNRAS.383.1000D}.}
\label{tab:1}       
%
%
\begin{tabular}{p{2cm}p{2.4cm}p{2cm}p{4.9cm}}
\hline\noalign{\smallskip}
Galaxy & age range & $\zeta$ & Reference  \\
\noalign{\smallskip}\svhline\noalign{\smallskip}

SMC & $20-1000$~Myr & $0.0\pm0.1$$^c$ & \cite{2007ApJ...668..268G}\\
M31 & $5-100$~Myr & $0-  0.15$ & Fouesneau et al.~2014 \\
NGC~2997 & $10-100$~Myr & $0.1\pm0.2$ & Ryon et al.~2014 \\
M51 & $10-300$~Myr & $0.15\pm0.2$ & Hwang \& Lee~2010 \\
Solar neighbourhood & $5-300$~Myr & $0.3\pm0.15$& Lamers et al.~2005\\
LMC & $10-100$~Myr & $0.3\pm0.15$ & Baumgardt et al.~2013 \\
M33 & $10-100$~Myr & $0.3\pm0.2$$^a$ & Gieles \& Bastian~2008$^b$ \\
NGC~4041 & $5-200$~Myr & $0.4\pm0.2$& \cite{2013AJ....145..137K}\\
NGC~1566 & $5-300$~Myr & $0.5\pm0.15$ & Hollyhead et al.~in prep.\\
NGC~4449 & $5-500$~Myr & $0.5\pm0.15$$^a$ & Annaballi et al.~2011 \\
NGC~7793 &$10-500$~Myr & $0.55\pm0.2$ & Silva-Villa \& Larsen~2011\\
NGC~1313 & $10-500$~Myr & $0.6\pm0.1$ & Silva-Villa \& Larsen~2011\\
\hline
M83 &$10-500$~Myr & $0.25\pm0.1$ & Silva-Villa \& Larsen~2011 \\
M83 F1 &$1-1000$~Myr&$0.9\pm0.2$ & Chandar et al.~2010b\\
M83 F2 & $10-1000$~Myr & $0.5\pm0.2$ & Chandar et al.~2014\\
M83 F2 & $5-300$~Myr & $0.15\pm0.15$ & Chandar et al.~2014 catalogue\\
M83 (F1-F7) & $10-300$~Myr & $0-0.6$&Silva-Villa et al.~2014\\
M83 (Full sample) & $10-300$~Myr & $0.35\pm0.15$&Silva-Villa et al.~2014\\
\hline
Antennae & $5-500$~Myr & $0.85\pm0.15$ & Whitmore et al.~2007, 2010\\
\noalign{\smallskip}\hline\noalign{\smallskip}
\end{tabular}
\end{table}


%

\section{Conclusions and Future Outlook}

Recent work on cluster populations has found an increasing level of connectedness between the population properties and those of the host galaxy.  It appears that the fraction of star-formation that happens in bound clusters ($\Gamma$) increases with the surface density of star-formation (which is likely just a proxy for the surface density of dense gas within a galaxy), ranging from $5-10$\% for quiescent spirals and dwarf galaxies to $\sim30-50$\% in starbursts.  Even within a single galaxy, $\Gamma$ can vary by a factor of four or more.  An interesting implication of this, is that star-formation in clusters may have been much more common in the early Universe, during the epoch of globular cluster formation.  While the cluster initial mass function is well described by a power-law with index $-2$ over much of the observed range, an increasing number of studies have found that there is a truncation (or break) at high-masses, the point of which, $\mstar$, is also dependent on the host galaxy properties.  Recent theoretical work (Kruijssen~2014) has linked $\mstar$ with the mass of the most massive GMCs within a galaxy (controlled by the Toomre-mass), hence galaxies like merging gas-rich systems which can produce massive GMC complexes are able to form more massive clusters, hence have higher values of $\mstar$.  Finally, cluster populations have been used to study the process of cluster disruption, with the age distribution of clusters being sensitive to the rate at which clusters are destroyed.  A clear trend of the age distribution with galaxy properties (with gas-rich high mass galaxies having steep age distributions, and quiescent galaxies having flat distributions) has been found by a number of studies.  For most galaxies, rapid disruption of young ($<100$~Myr) clusters is not supported by the data, and that the lifetimes of clusters are strongly related to their ambient environment.

Many of the studies and results presented here are based on a limited number of observations, or a small sample of cluster populations.  Hence, large surveys focussing on individual galaxies (such as the M31 {\em Panchromatic Hubble Andromeda Treasury} (PHAT) survey - Dalcanton et al.~2012) or large galaxy samples (e.g., Legacy Extragalactic UV Survey - Calzetti et al.~2015) will allow detailed tests of the relations presented here as well as our theoretical framework to understand them.  How does $\Gamma$ vary within galaxies, both in space and time?  Are there environments where cluster formation is actively suppressed?  Or environments that encourage the formation of only a handful of massive clusters instead of sampling from an underlying parent distribution that favours the formation of many low-mass clusters (i.e., like that observed in most galaxies)?

On the cluster disruption side, the influence of environment on the lifetime of clusters is clear, and is expected for all scenarios of cluster dissolution.  However, the role of cluster mass is still uncertain.  The dominant cluster disruption mechanism in many galaxies is interactions with passing GMCs.  The effect of the passage is proportional to the cluster density, hence if there is not a specific mass-radius relation for young clusters, cluster disruption is expected to be dependent on cluster mass (with high mass clusters surviving longer).  Hence, deriving the cluster mass-radius relation in a sample of galaxies, and at a range of ages, will be very useful.  Looking for changes in the cluster mass function (at the low mass end) as a function of age within a population is also a potential way to estimate the dependence of mass on disruption, however, incompleteness and sample selection affect the low-mass end of any observed sample preferentially, making the distinction between selection effects and physical properties challenging.  Larger samples of cluster populations, however, may be able to address this question statistically.

Finally, one of the outstanding questions of cluster research is how, exactly, do the young massive clusters observed today relate to the ancient globular clusters observed around all major galaxies.  Can we simply apply our understanding of cluster formation locally, and scale to the conditions of the early Universe?  Much theoretical progress has been made in linking globulars and young massive clusters (e.g., Kravtsov \& Gnedin~2005; Kruijssen 2014), however many open issues remain. 

%
%
%
%
\begin{acknowledgement}
We gratefully acknowledge comments and suggestions by Diederik Kruijssen, S{\o}ren Larsen, Linda Smith, Steven Stahler, Jenna Ryon, Esteban Silva-Villa, Iraklis Konstantopoulos, Mark Gieles, and Daniella Calzetti which significantly improved the manuscript.  NB is partially funded by a Royal Society University Research Fellowship.
\end{acknowledgement}
%
%
%


\input{reference_v1_nb_aa2}

\end{document}

%% file: reference_v1_nb_aa2.tex
%
%
%